\documentclass[11pt]{article}
\usepackage{graphicx}
\usepackage{amsmath}
\usepackage{graphicx}
\usepackage{amsfonts}
\usepackage{amssymb}
\usepackage{bbm}
\usepackage[colorlinks]{hyperref}
\usepackage{enumitem}
\usepackage{framed}
\usepackage{caption}
\usepackage{subcaption}
\usepackage{epstopdf}
\usepackage[top=3cm,bottom=3cm,left=3cm,right=3cm]{geometry}
\usepackage{url}
\usepackage{booktabs}
\newcommand{\m}{\mathbb}
\newcommand{\n}{\nonumber}
\date{}

\begin{document}
\title{Skewing Quanto with Simplicity}
\author{George Hong}

\maketitle
\begin{center}
\vspace{-.5cm}
Credit Suisse \\
\texttt{george.hong@credit-suisse.com} \\
\vspace{.5cm}
First Version: 8-Jul-2018 \\
This Version: 25-Mar-2019
\end{center}

\begin{abstract}
We present a simple and highly efficient analytical method for solving the Quanto Skew problem in Equities under a framework that accommodates both Equity and FX volatility skew consistently. Ease of implementation and extremely fast performance of this new approach should benefit a wide spectrum of market participants.
\end{abstract}

\section{Introduction}
\label{sec:Introduction}
Quanto derivatives have existed for almost as long as derivatives quants. They are particularly popular in the Equities OTC and structured product markets where a significant portion of the trades globally are quantos. It has become such a common and fundamental product feature nowadays that people often take the market participants' ability to properly price and risk manage quantos for granted. In the Black-Scholes model, an elegant covariance drift adjustment enables one to accurately value and compute risks of a quanto derivative \emph{without increasing the dimensionality} of the problem. This insight is applicable to valuing general payoffs in the absence of volatility skew and have therefore become not just an integral part of many production pricing libraries but also the basis for model intuition for quants and traders alike: the hedging cost for the cross gamma due to the quanto feature is captured by the covariance between the underlying and the FX assets in the drift.

It is in fact common for practitioners to retain the simplified approach, even when volatility skew is assumed elsewhere for equity, rates or other risk factors in a model (with the exception of the FX asset class where the vol skews for all the triangular pairs are too obvious to be ignored). By keeping the drift term of the asset dynamics constant and devising some ad-hoc rules on picking \emph{some} volatility level, one could potentially model the volatility skew in the \emph{diffusion} part only, while staying well-marked on the tradable quanto forward prices using an implied correlation as an fudge factor. Although simple and fast, this approach has significant drawbacks, as explained in \cite{Jackel} and \cite{Vong}.

The key reasons for the popularity of the "ad-hoc" adjustment approach lie in its simplicity, the (perceived) universality, low implementation effort and zero incremental compute costs. Hence it would be strongly desirable to address its deficiencies while maintaining its advantages where we can. With those goals in mind, we apply the Stochastic Collocation method recently introduced to finance from the field of Uncertainty Quantification by \cite{Grzelaka, Grzelaka2}, where the authors developed novel techniques to speed up Monte Carlo sampling and remove vol arbitrages.

Our contribution lies in the formulation of a simple and highly efficient recipe for computing quantities related to quanto valuations using the stochastic collocation techniques. The method can be used to quickly compute quanto option prices, calibrate to observed quanto forward for an implied correlation and determine the local drift function in the sense of Markovian projection as defined in \cite{RenMadanQian}. The first two results allow a wide audience of market participants to price European quanto options with minimal implementation effort and no reliance on time-consuming numerical methods. The last result, on the other hand, enables modellers to incrementally improve their quanto methodology without incurring additional costs to their risk management systems and quoting platforms.

Recently several other authors have also proposed novel approaches to overcome the extra compute and speed requirement. For example, \cite{Vong} and \cite{Hok} have applied perturbation methods to obtain so-called \emph{Proxy expansion} formulae for European quanto option prices. This paper joins the quest for an efficient methodology that bypass PDE or Monte Carlo methods. However, our approach differs in several aspects with the key difference being that the perturbation approach is "local" in nature as the proxy expansion is centered around \emph{one} point only (often the ATM spot or the option strike). This means one could have two input volatility surfaces that differ materially on the wings but agree well locally (around the current spot, say) in terms of their low-order derivatives and the Taylor-like expansion formulae would produce very similar quanto option prices. On the other hand, the \emph{Polynomial expansion} approach we proposed below is "global" in nature as it makes use of volatilities across a wide strike range via Lagrange interpolation. Finally, in contrast to other research employing methods such as Fast Fourier Transform within the Affine diffusion framework, no specialised assumption is made here on the underlying processes.

\section{Quantos and Quantiles}
\label{sec:QuantosQuantiles}
\subsection{Quanto Basics}
We follow the general convention and call the equity asset foreign and its base currency the \emph{Foreign} currency. The payoff currency, which the structure is "quantoed into", is called the \emph{Domestic} currency. The FX rate, $ X_t $, is the number of unit of Foreign currency per 1 unit of Domestic currency. Consider the risk-neutral dynamics in the foreign measure, $\m F$:
\begin{eqnarray}
\label{eq:dynamicsF}
   \tfrac{d S}{S} &=&  (r_{_F}-\delta)dt + \sigma_{_S}(S_t, t) \; d W_{^S}^{\m F} \n \\
   \tfrac{d X}{X} &=&  (r_{_F}-r_{_D})dt + \sigma_{_X}(X_t, t) \; d W_{^X}^{\m F} \;\; . \n
\end{eqnarray}
where $r, \delta, \sigma(\cdot, \cdot)$ denote the deterministic risk-free rates, equity dividend yield and local volatility functions, and $\m E_{\m F}[ d W_{^S}^{\m F} d W_{^X}^{\m F}]= \rho \, dt$. Note that the market reality remains that the only liquid instrument with which one can calibrate to and hedge the Equity-FX covariance is currently still limited to Quanto forwards or futures. This implies that the modelling of correlation skew is perhaps less pressing than say, in the FX world where the volatility surface of the "Cross" also needs to be repriced as in \cite{BennettKennedy}. Hence a model with constant correlation focusing on the volatility skews serves as a good step forward that provides a consistent framework for valuing equity quantos.

Under the Domestic measure, $\m D$, we have:
\begin{eqnarray}
\label{eq:dynamicsD}
   \tfrac{d S}{S} &=&  (r_{_F}-\delta + \rho \, \sigma_{_X} (X_t, t) \sigma_{_S} (S_t, t) )dt + \sigma_{_S}(S_t, t) \; d W_{^S}^{\m D} \n \\
   \tfrac{d X}{X} &=&  (r_{_F}-r_{_D}+\sigma_{_X}(X_t, t)^2)dt + \sigma_{_X}(X_t, t) \; d W_{^X}^{\m D} \;\; . \n
\end{eqnarray}

Consider valuing Quanto, vanilla equity and FX calls under Domestic and Foreign measures:
\begin{align*}
   C^Q({K,T}) &= B_{_D}^{_T} \cdot  \m E_{\m D} \Big[ \big( S_{_T}-K \big)_+ \Big]= B_{_F}^{_T} \cdot  \m E_{\m F} \Big[  \tfrac{X_{_T}}{X_0}\cdot \big( S_{_T}-K \big)_+ \Big], \\
   C^S({K,T}) &= B_{_D}^{_T} \cdot \m E_{\m D} \Big[ \tfrac{X_{_0}}{X_{_T}}  (S_{_T}-K)_+ \Big]= B_{_F}^{_T} \cdot \m E_{\m F} \Big[ \big( S_{_T}- K \big)_+ \Big], \\
   C^X({K,T}) &= B_{_D}^{_T} \cdot \m E_{\m D} \Big[ \tfrac{X_{_0}}{X_{_T}}  (X_{_T}-K)_+ \Big]= B_{_F}^{_T} \cdot \m E_{\m F} \Big[ \big( X_{_T}- K \big)_+ \Big],
\end{align*}
with Domestic/Foreign discount factors as $B_{_D}^{_T} := e^{(-r_{_D} T)}; \;\;\; B_{_F}^{_T} := e^{(-r_{_F} T)}.$

For any general European payoff paying $V_{_T}(S_{_T}, X_{_T})$ at time $T$, its price in domestic currency is:
\begin{eqnarray*}
   V_{0} = B_{_D}^{_T} \cdot \m E_{\m D} \Big[ V_{_T} \Big] = B_{_F}^{_T} \cdot \m E_{\m F} \Big[ \tfrac{X_{_T}}{X_0} \cdot V_{_T} \Big], \;\;\;\;\; \frac{d \m D}{d \m F} = \frac{X{_T}}{X{_0}} \frac{B_{_F}^{_T}}{B_{_D}^{_T}} = \frac{X{_T}}{\m E_{\m F} \big[ X_{_T} \big]}.
\end{eqnarray*}

\subsection{Quantile Transform}
We briefly recall the well-known technique of distribution mapping widely used on the street (a.k.a. quantile transform). Given two random variables, $Y$ and $Z$, whose CDF, $F_{_Y}(y) := \m P(Y<y)$ and $F_{_Z}(z) := \m P(Z<z)$ are monotonic and right-continuous, there exist an inverse function, called \emph{quantile function}, in the following sense:
$$ F_{_Y}^{-1}(q) = \inf \{ y | F_Y(y) \geq q, 0<q<1 \}.
$$
If we define the random variable $Y':=g(Z)$ with the \emph{Quantile transform} function: $g(z) := F_{_Y}^{-1}({F_{_Z}}(z))$, then the fact that the CDF of a random variable is uniformly distributed implies that $Y'$ has the same distribution as $Y$.

Now let $F_{X_{_T}}$ and $F_{S_{_T}}$ be the CDF's of the FX and Equity underlyings at maturity $T$ in the foreign measure, $\m F$. They are determined by the full Vanilla European call prices, $C^{^X}(K,T)$ and $C^{^S}(K,T)$:
\begin{eqnarray}
\label{eq:cdfCallPrices}
  F_{X_{_T}}(K) &=& 1+e^{(r_{_F} T)} \cdot \tfrac{\partial C^{X}(K,T)}{\partial K}  \;  \n \\
  F_{S_{_T}}(K) &=& 1+e^{(r_{_F} T)} \cdot \tfrac{\partial C^{S}(K,T)}{\partial K}  \;
\end{eqnarray}
which can be equivalently re-expressed in terms of the calibrated implied volatility surface, $\hat{\sigma}^{X}(K,T)$ and $\hat{\sigma}^{S}(K,T)$. Care needs to be taken in the tail extrapolation of the implied vol to prevent arbitrages and ensure accuracy and stability (see \cite{Cesarini} for detailed discussions). This leads to the following distribution maps for the underlyings:
\begin{eqnarray}
\label{eq:distMap}
  g_1(z) = F^{-1}_{X_{_T}}(N(z)), \;\;\; g_2(z) = F^{-1}_{S_{_T}}(N(z))
\end{eqnarray}
where we map the $\m F$-distributions of $X_{_T}$ and $S_{_T}$ onto the standard Normal distribution. So working with a standard Normal $Z$ will give us $g_1(Z)$ that has $F_{X_{_T}}$ as the distribution function.

\subsection{Stochastic Collocation}
We briefly introduce the concept and refer the readers to \cite{Grzelaka} and the references therein for a general introduction to the field of Polynomial Chaos. Consider the Lagrange polynomial approximation for a general function $u(\cdot)$:
\begin{eqnarray*}
  u(x) \approx \tilde{u} (x) := \sum_{i=0}^{N} u(x_i) L_i (x), \; \;
  L_i(x) = \prod_{j=1,j \neq i}^{N} \frac{x-x_j}{x_i-x_j}
\end{eqnarray*}
where $\{x_i\}$ are the nodes and $L_i(x)$ are the Lagrange interpolation polynomials satisfying the orthogonal property $L_i(x_j) = \delta_{ij}$. We re-expressed this as:
$$ \tilde{u}(x) = a_0 + a_1x + \cdots + a_{N-1}x^{N-1} = \textbf{a}^{\top} \textbf{poly}_{(N-1)}(x),
$$
where we denote $\textbf{poly}_{(N-1)}(x) = (1,x,x^2,\cdots,x^{N-1})^{\top}$. The coefficients $\textbf{a} = (a_0, a_1,a_2,\cdots,a_{N-1})^{\top}$ solves the linear system with the Vandermonde matrix, V, below:
\begin{eqnarray}
\label{eq:vandermonde}
& V \textbf{a} = u(\textbf{x}) \\
\left(
\begin{array}{ccccc}
  1 & x_1 & x_1^2 &  .. & x_1^{N-1} \\
  1 & x_2 & x_2^2 &  .. & x_2^{N-1} \\
  \vdots &  \vdots  & \vdots &  .. & \vdots \\
  1 & x_{N} & x_{N}^2 &  .. & x_{N}^{N-1}      \end{array}  \right)
&
\left( \begin{array}{c} a_0 \\ a_1 \\ \vdots \\ a_{N-1} \end{array} \right)
=
\left( \begin{array}{c} u(x_1) \\ u(x_2) \\ \vdots \\ u(x_{N}) \end{array} \right) \;\; \n
\end{eqnarray}
Following \cite{Grzelaka} we let $u(\cdot)$ be the distribution mapping function \eqref{eq:distMap} and the nodes $\{x_i\} $ be the zeroes of the Gauss-Hermit polynomials (a natural choice given the mapping to Normal distributions). Solving the two set of coefficients $\textbf{a}_1$ and $\textbf{a}_2$ gives us polynomial approximations to the distribution mapping (e.g. to solve a small Vandermonde system see \cite{PressNumericalrecipes} for simple routines or simply do matrix inversion with double precision). By choosing the Bivariate Normal as the driving random variables, $Z_1, Z_2$, correlated by $\rho$ and the marginal distribution mappings, we now have a pair of transformed random variables with joint distribution in the Gaussian Copula setting (where we may choose the polynomials to have different orders):
\begin{eqnarray*}
  & X_{_T} \overset{d}{=} g_{1}(Z_1) \approx \tilde{g}_{1}(Z_1) := \sum_{n=0}^{N_1 - 1} a_{1,n} \cdot Z_1^n\\
  & S_{_T} \overset{d}{=} g_{2}(Z_2) \approx \tilde{g}_{2}(Z_2) := \sum_{n=0}^{N_2 - 1} a_{2,n} \cdot Z_2^n
\end{eqnarray*}
As articulated in \cite{Bergomi} on the resemblance between multi-underlying constant correlation Local Volatility models and a model with Gaussian copula linking the marginal distributions, this in turns leads us to the approximation of the $T$-distribution for the local vol processes of the Equity and FX assets by the method that follows.

\section{Quanto Skew Quantified}
\label{sec:QuantoSkewQuantified}
\subsection{Conditional Expectation as Polynomial Expansion}
First note the following result regarding the \emph{conditional} moments of a Gaussian on another correlated Gaussian:
\begin{eqnarray}
\label{eq:momentsConditionalNormal}
\m E [ Z_1^n  | Z_2 = z ] &=& \m E [ \bar Z_1(z) ^n ] \n = \sum_{j=0}^{ \lfloor \frac{n}{2} \rfloor } {n \choose 2 j}  (2 j -1 )!! (\sqrt{1-\rho^2})^{(2j)} (\rho z)^{(n-2j)} \n \\
&=:& \sum_{i=0}^{n} q_i(n;\rho) \cdot z^i = {\textbf{q}(n;\rho)}^{\top} \; \textbf{poly}_n(z)
\end{eqnarray}
which comes from the conditioned variable again being Normally distributed: $ \bar Z_1 \sim \mathcal{N}(\rho z, \sqrt{1-\rho^2}) $ (and $n!!$ is the double factorial function which multiplies all integers up to $n$ with the same parity). The conditional moments are polynomials with coefficients $q_i(n;\rho)$ dependent on the correlation.

Next we compute the conditional expectation of FX price given the Equity \emph{Gaussian driver}, $Z_2$:
\begin{eqnarray}
\label{eq:xConditionalOnZ2}
  && \m E_{\m F} [ X_{_T} | Z_2 ] \approx \m E_{\m F} \Big [ \tilde{g}_1(Z_1) \Big | Z_2 \Big ] = \sum_{n=0}^{N_1 - 1} a_{1,n} \cdot \m E_{\m F} \Big [ Z_1^n \Big | Z_2 \Big ] \n \\
   &&=  \sum_{n=0}^{N_1 - 1} a_{1,n} \Big( \sum_{j=0}^{n} q_j Z_2^j \Big) = \sum_{n=0}^{N_1 - 1} b_{n} Z_2^n = \textbf{b}^{\top} \; \textbf{poly}_{\scriptscriptstyle{(N_1-1)}}(Z_2)  \;\;\;\;\; \n \\
  && b_{n} :=   \sum_{j=n}^{N_1 - 1} a_{1,j} \; q_n(j; \rho)
\end{eqnarray}
The key is to approximate the conditional expectation of $X$ as a polynomial where the new coefficients $\textbf{b}$ are scaled from the original coefficients $\textbf{a}_1$ by a multiplier calculated from the conditional moments of the Bivariate Normal distribution, allowing an expression as a function of the Equity price:
$$ \m E_{\m F} [ X_{_T} | S_{_T}=S ] = \m E_{\m F} [ X_{_T} | Z_2=g_2^{-1}(S) ] \approx \sum_{n=0}^{N_1 - 1} b_{n} \cdot \big( g_2^{-1}(S) \big)^n
$$
When both $X_{_T}$ and $S_{_T}$ are Log-Normal, this conditional expectation is proportional to a simple power function of S where the power is the correlation times the ratio of the volatilities. Introducing Equity and FX skews complicates the function shape, adding more convexities and turns. This is illustrated here by a graph of the conditional expectation function using 2017 market skew for Nikkei and USDJPY alongside the "No Skew" case with Log-Normal distribution given by ATM vols. We see that our successive polynomial approximation with increasing order quickly converges to the exact function solved by a 2-d integration.
\begin{figure}[!htb]
\centering
\includegraphics[width=3.0in]{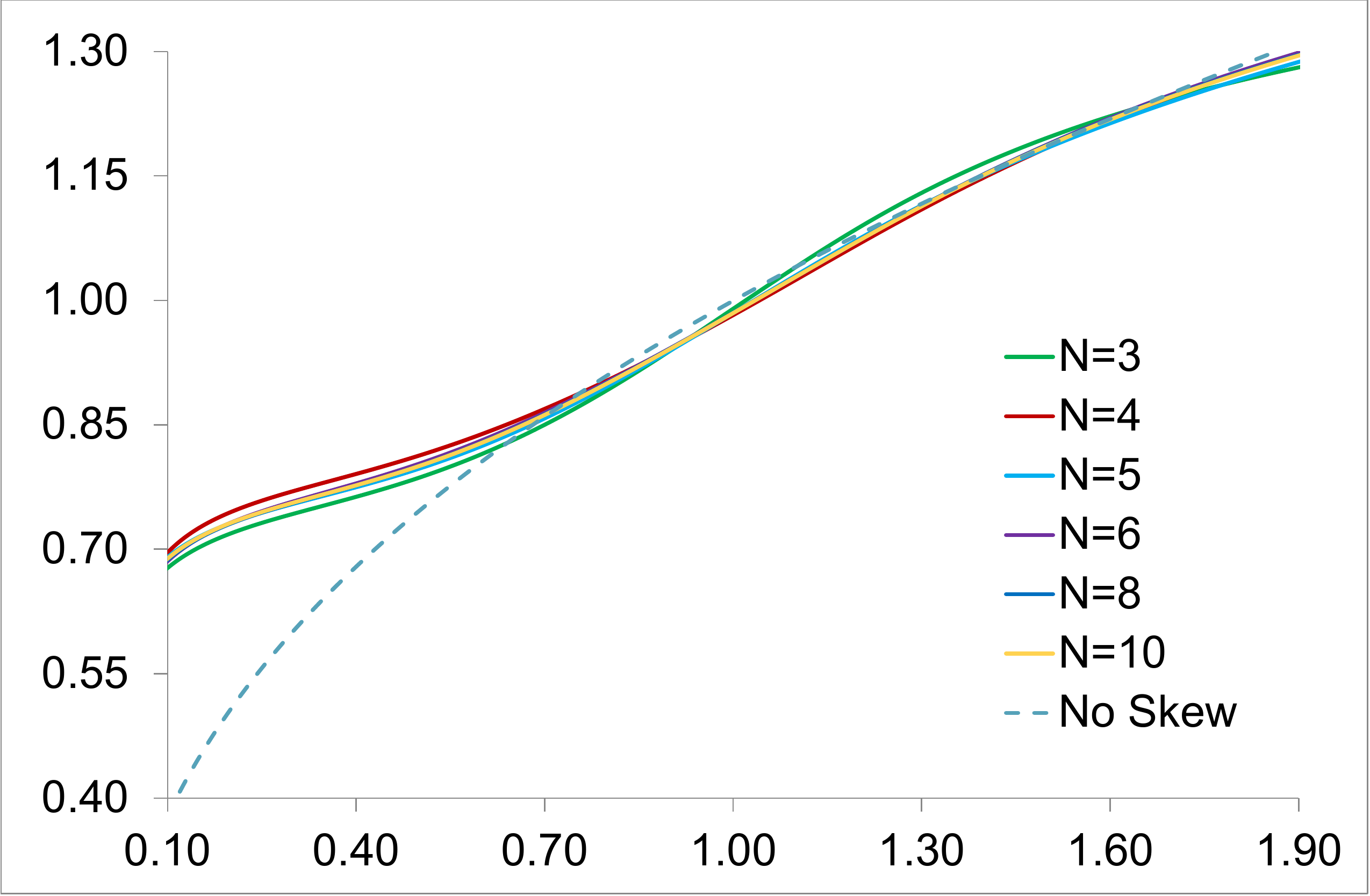}
\captionsetup{singlelinecheck=off}
\caption[.]{Approximation of conditional expectation $\m E_{\m F} [\frac{X_{_T}}{X_{0}} | S_{_T} ] $ as a function of $ \frac{S_{_T}}{S_{0}} $}
\end{figure}
\subsection{Quanto Vanilla Spread}
\label{subsec:QuantoVanillaSpread}
We now value the Quantos. Given the potential oscillatory nature of polynomial approximation impacting the accuracy of the analytical method, we use the non-quanto European call (of the same strike) as a control variate for "noise cancelation". To simplify notation, we work with undiscounted option values scaled by initial FX, $ \hat C_{_{T,K}}^Q : = C_{_{T,K}}^Q / B_{_F}^{_T} \cdot X_0 $ and $ \hat C_{_{T,K}}^S : = C_{_{T,K}}^S / B_{_F}^{_T} \cdot  X_0 $ and consider the equivalent strike in the Gaussian space by mapping $K$ via the inverse of the distribution mapping $g_2(\cdot)$: $ \kappa := g_2^{-1}(K). $

The idea is to re-express in terms of the driving Gaussians and the conditional moments \eqref{eq:xConditionalOnZ2}:
\begin{eqnarray}
\label{eq:quantoCallExpansion}
   \hat C_{_{T,K}}^Q &=& \m E_{\m F} \Big[  X_{_T}  \big( S_{_T}-K \big)_+ \Big] \n \approx \m E_{\m F} \Big[ \widetilde{g_1}(Z_1)  \cdot \big( \widetilde{g_2}(Z_2)-K \big) \mathbbm{1}_{\{ g_2(Z_2) >K \}} \Big] \n \\
   &=&  \m E_{\m F} \Big[ \m E_{\m F} [ \widetilde{g_1}(Z_1) | Z_2] \cdot \big( \widetilde{g_2}(Z_2)-K \big) \cdot \mathbbm{1}_{\{ g_2(Z_2) >K \}} \Big] \n \\
   &=&  \m E_{\m F} \bigg[  \sum_{i=0}^{N_1 - 1} b_{i} Z_2^i \cdot \bigg( \sum_{j=0}^{N_2 - 1} a_{2,j} Z_2^j -K \bigg) \cdot \mathbbm{1}_{\{ Z_2 > \kappa \}} \bigg] \n \\
   &=&  \m E_{\m F} \bigg[ \bigg( \sum_{n=0}^{N_1 + N_2 - 2} c_{n} Z_2^n - K \cdot \sum_{n=0}^{N_1 - 1} b_{n} Z_2^n \bigg) \cdot \mathbbm{1}_{\{ Z_2 > \kappa \}} \bigg] \\
\label{eq:convolveConditionalAndS}
  c_n &:=& \sum_{k=0}^{n} a_{2,k}  b_{n-k}, \; \; \; \forall \; n = 0, \cdots, N_1+N_2-2
\end{eqnarray}
Note the vector $\textbf{c}$ is the convolution of $\textbf{a}_2$ and $\textbf{b}$ which makes intuitive sense as it defines the polynomial coefficients for $X_{_T} \times S_{_T}$ via conditioning. Let's denote $N-1:= N_1+N_2-2$ so the total number of coefficients to keep track is at most $N$.

Now following \cite{Grzelaka} we make use the recursive relation for the moment formulae for truncated normal distribution:
\begin{eqnarray}
\label{eq:momentsRecursive}
   m_i(\kappa) &:=& \m E [  Z^i  | Z > \kappa ], \; \; \; \; Z \sim N(0,1), \; \;   \n  \\
               &=& (i-1) \cdot m_{i-2}(\kappa) + \kappa^{i-1} \frac{\phi(\kappa)}{1-\Phi(\kappa)} \; \; \\
   m_0(\kappa) &=& 1, \; \; m_{-1}(\kappa) = 0 \n
\end{eqnarray}
where the Normal CDF $\Phi(\cdot)$ and PDF $\phi(\cdot)$ are evaluated once only regardless of $N$.

This allows us to compute each term in the formula above:
\begin{eqnarray}
\label{eq:momentsTruncatedNormal}
   \m E \Big [  Z_2^i   \cdot \mathbbm{1}_{\{ Z_2 > \kappa \}}  \Big ] = \m E \big [  Z_2^i  \big | Z_2 > \kappa \big ] \cdot \m P \big [ Z_2 > \kappa \big ]
   = m_i ( \kappa ) \cdot \big[1- \Phi(\kappa) \big]
\end{eqnarray}
For the (Foreign) Equity Vanilla Call, we have a similar formula in the coefficient $\{a_{2,n}\}_0^{N_2-1}$:
\begin{eqnarray}
\label{eq:vanillaCallExpansion}
   \hat C_{_{T,K}}^S \approx X_0 \cdot \m E_{\m F} \Big[ \big( \widetilde{g_2}(Z_2)-K \big)_+ \Big] = X_0 \cdot \m E_{\m F} \bigg[ \bigg( \sum_{n=0}^{\scriptscriptstyle N_2 - 1} a_{2,n} Z_2^n - K \bigg) \cdot \mathbbm{1}_{\{ Z_2 > g_2^{-1}(K) \}} \bigg]
\end{eqnarray}

Combining \eqref{eq:quantoCallExpansion} and \eqref{eq:vanillaCallExpansion}, we get the Quanto-Vanilla Spread with strike $K$ as a polynomial expansion:
\begin{eqnarray}
\label{eq:quantoVanillaSpreadExpansion0}
   & & \hat C_{_{T,K}}^{QS} := \hat C_{_{T,K}}^Q - \hat C_{_{T,K}}^S = \m E_{\m F} \Big[  X_{_T}  \big( S_{_T}-K \big)_+ \Big] - X_0 \m E_{\m F} \Big[ \big( S_{_T}-K \big)_+ \Big] \n \\
   & & =  \m E_{\m F} \bigg[ \bigg( \sum_{n=0}^{N-1} c_{n} Z_2^n - K \cdot \sum_{n=0}^{N_1 - 1} b_{n} Z_2^n - X_0 \cdot \sum_{n=0}^{N_2 - 1} a_{2,n} Z_2^n + X_0 \cdot K \bigg) \cdot \mathbbm{1}_{\{ Z_2 > \kappa \}}  \bigg] \\
   & & =  \m E_{\m F} \bigg[ \bigg( \sum_{n=0}^{N - 1} e_{n}(K) Z_2^n \bigg) \cdot \mathbbm{1}_{\{ Z_2 > \kappa \}} \n \bigg] = \sum_{n=0}^{N - 1} e_{n}(K) \cdot \m E_{\m F} \bigg[  Z_2^n \cdot \mathbbm{1}_{\{ Z_2 > \kappa \}} \n \bigg] \n \\
\label{eq:quantoVanillaSpreadExpansion}
   && =  \bigg( \sum_{n=0}^{N - 1} e_{n}(K) \cdot m_n ( \kappa ) \bigg) \cdot \Big[1- F_{S_{_T}}(K) \Big]
\end{eqnarray}
where we have collected the coefficients $\textbf{c}, \textbf{b}, \textbf{a}_2$ in \eqref{eq:quantoVanillaSpreadExpansion0} into $\textbf{e}$:
\begin{eqnarray}
\label{eq:eDefinitionFull}
\left(
    \begin{array}{c} e_0 \\ \vdots \\ e_{(N_1-1)} \\ . \\ e_{(N_2-1)} \\ \vdots \\ e_{(N-1)}
    \end{array}
\right) \; \; :=
\left(
    \begin{array}{c} c_0 \\ \vdots \\ c_{(N_1-1)} \\ . \\ c_{(N_2-1)} \\ \vdots \\ c_{(N-1)}
    \end{array}
\right) \;\; - K \cdot
\left(
    \begin{array}{c} (b_0-X_0) \\ \vdots \\ b_{(N_1-1)} \\ . \\ 0 \\ \vdots \\ 0
    \end{array}
\right) \;\; - X_0 \cdot
\left(
    \begin{array}{c} a_{2,0} \\ \vdots \\ \vdots \\ . \\ a_{2,{(N_2-1)}} \\ \vdots \\ 0
    \end{array}
\right) \;\; \n
\end{eqnarray}
\begin{eqnarray}
\label{eq:eDefinition}
 e_{n}:= c_{n} - K \cdot b_{n} \mathbbm{1}_{\scriptscriptstyle n<N_1} - X_0 \cdot a_{2,n} \mathbbm{1}_{\scriptscriptstyle n<N_2} + X_0 K \mathbbm{1}_{\scriptscriptstyle n=0}
\end{eqnarray}

We now summarize the pricing recipe as:
\begin{framed}
\label{methodOption}
\begin{enumerate}[label=(\roman*)]
\item Choose $N_1, N_2$ and evaluate $g_1, g_2$ on $\{x_i\}$ per \eqref{eq:cdfCallPrices}
\item Solve the Vandermonde system in \eqref{eq:vandermonde} for coefficients $\textbf{a}_1$, $\textbf{a}_2$;
\item Calculate the conditional coefficients $\textbf{b}$ with \eqref{eq:xConditionalOnZ2} by summing $q_j$ in  \eqref{eq:momentsConditionalNormal}. Calculate $\textbf{c}$ from $\textbf{a}_2$ and $\textbf{b}$ with \eqref{eq:convolveConditionalAndS}
\item Given strike $K$, calculate $\kappa$ and then $\textbf{m}$ with \eqref{eq:momentsRecursive}.
\item Sum-product $\textbf{m}$ with $\textbf{e}$ in \eqref{eq:eDefinition}, scale by CDF for Quanto-Vanilla Spread in \eqref{eq:quantoVanillaSpreadExpansion}.
\end{enumerate}
\end{framed}
The calculated Quanto-Vanilla Spread value, $C^{QS}$, can now be added to the available Vanilla Call price, $C^{S}$, to get the Quanto option price, $C^{Q}$.

The method is very fast (see results in Table \ref{tab:computeTime}) because all the steps involve elementary arithmetic/matrix operations only. In addition, results from several steps are identical for options with the same maturity but different strikes. For example, the coefficients from Step (i), (ii) and (iii) ($\textbf{a}_1, \textbf{a}_2, \textbf{b}, \textbf{c}$) are all independent of $K$ and need not be recalculated: we only need to update $\textbf{m}$ (evaluating the Normal CDF and PDF \emph{once} only per one strike) and $\textbf{e}$ (which is trivial).

\subsection{Quanto Local Drift Adjustment}
Quantos exist in many derivative products beyond European options. To accurately capture the FX/Equity skews in the quanto drift the most accurate way is to evolve an additional FX process (either in Monte Carlo simulation or PDE grid), effectively treating the Quanto product as a FX hybrid with an additional stochastic factor. While it is perfectly possible to implement this approach in the production pricing libraries to support all payoffs, the additional computational costs and implementation efforts mean that it is a common practice for industry models to rely on approximations and avoid increasing the dimensionality of the valuation problem for every single quanto trade.

Here we follow \cite{RenMadanQian} and apply the celebrated Gy\"{o}ngy projection approach \cite{Gyongy} to the 2-d $\m D$ dynamics for an 1-d Markov process. This ensures all the Quanto European options are correctly re-priced across maturities.
\begin{eqnarray}
\label{eq:GyongyQuanto}
  \frac{d S}{S} &=&  \bigg( r_{_F}-\delta + \m E_{\m D} \Big[ \rho \, \sigma_{_S}(S_t, t) \, \sigma_{_X}(X_t, t) \Big | S_t \Big] \bigg) \, dt + \sigma_{_S}(S_t, t) \; d W_{^S}^{\m D}  \n \\
  &=& \Big(r_{_F}-\delta + \rho \; \sigma_{_S}(S_t, t) \; \sigma_{_{XS}}(S_t, t) \Big) \, dt + \sigma_{_S}(S_t, t) \; d W_{^S}^{\m D} \n
\end{eqnarray}
with the new function in the drift defined as:
\begin{eqnarray}
\label{eq:conditionalLVDomestic}
  \sigma_{_{XS}}(S, t) &:=& \m E_{\m D} \big [\sigma_{X}(X_t, t) \big | S_t = S \big ]
\end{eqnarray}
The authors in \cite{RenMadanQian} shared the non-trivial result that the FX local volatility transfers into a localised drift in the one-dimensional law for the quantoed stock, akin to the well-known quanto adjustments. However they also noted that "there is no more direct way of calculating the adjustment". We will continue from that insight and calculate this adjustment using our new techniques, effectively obtaining a competitive way to compute this \emph{conditional Local Volatility} function of $X$ on $S$, $\sigma_{_{XS}}(S, t)$ and consequently the \emph{Quanto Local Drift} function, $ \rho \; \sigma_{XS}(S, t) \; \sigma_{S}(S, t) $.

Noting the expression \eqref{eq:conditionalLVDomestic} is under Domestic measure but all the implied volatility information for $\{S_t\}$ is under Foreign measure, we move to $\m F$:
\begin{eqnarray}
\label{eq:conditionalLVForeign}
  \sigma_{XS}(S, t) &=&  \m E_{\m F} \Big [\tfrac{X{_t}}{\m E_{\m F} [  X_{_t} ]} \cdot \sigma_{X}(X_t, t) \Big | S_t = S \Big ]
\end{eqnarray}
The key idea is to approximate the well-behaved forex local volatility $\sigma_{X}(X_t, t)$ again with the Lagrange polynomial. Mapping each $t$ slice of the distribution of $X_t$ to normals $Z_1$ with $g_1(\cdot)$ (dropping $t$ assumed implicitly in the notation), we have:
\begin{eqnarray}
\label{eq:nuDefinition}
  \nu_t(Z_1) &:=&  \Big( \tfrac{g_1(Z_1)}{X_0} \cdot \sigma_{X}( g_1(Z_1) , t) \cdot \tfrac{B_{_F}^{_T}}{B_{_D}^{_T}} \Big) \;\; \overset{d}{=}  \tfrac{X{_t}}{\m E_{\m F} [  X_{_t} ]} \cdot \sigma_{X}(X_t, t) \n \\
  &\approx& \tilde{\nu}_t(Z_1)  := \sum_{n=0}^{N_t - 1} a_{t, n} \cdot Z_1^n \n
\end{eqnarray}
The coefficients $\textbf{a}_t = (a_0, a_1,a_2,\cdots,a_{N_t-1})^{\top}$ again solves the linear equation $V \textbf{a}_t = \nu_t(\textbf{x})$ where $\textbf{x}$ are the Gaussian quadrature points which the re-mapped local vol functions $\nu(z)$ are applied. In practice we use the same $N_t$ for all $t$.

Note that we do \emph{not} need to construct the full FX local vol grid here. For each $t$ we only need to evaluate the $\nu(z)$ function $N_t$ times where $N_t$ is typically small ($ \leq 10$), which can be efficiently done using the standard local vol formula from the arbitrage-free vol surface (\cite{Gatheral}, \cite{Bergomi}). Similar to \eqref{eq:momentsConditionalNormal} and \eqref{eq:xConditionalOnZ2}, we now have:
\begin{eqnarray}
\label{eq:conditionalLVExpansion}
   \sigma_{XS}(S, t) &\approx&   \m E_{\m F} \bigg [\sum_{n=0}^{N_t - 1} a_{t, n} \cdot Z_1^n \bigg | Z_2 = g_2^{-1}(S) \bigg ] =  \sum_{n=0}^{N_t - 1} a_{t, n} \cdot  \m E_{\m F} \big [Z_1^n \big | Z_2 = z_{_S} \big ] \n \\
   &=&  \sum_{n=0}^{N_t - 1} a_{t, n} \cdot \Big( \sum_{i=0}^{n} q_i(n;\rho) \cdot (z_{_S})^i \Big) = \sum_{n=0}^{N_t - 1} b_{t, n} \cdot (z_{_S})^n  \n \\
   b_{t,n} &:=& \sum_{j=n}^{N_t - 1} a_{t,j} \; q_n(j; \rho) \n
\end{eqnarray}
where the re-mapped variable $z_{_S} : =g_2^{-1}(S) $ is cheap to evaluate.

With the Quanto Local Drift function at hand, we can now replace the ad-hoc constant quanto drift adjustments. Note that for strongly path-dependent payoffs, the local drift approach will not yield identical results as the full-blown 2-factor local volatility model in Equity and FX because Markovian projection guarantees the invariance in the terminal distribution only. On the other hand, for products such as basket, dispersion or rainbow options, the approach performs very well and allows easy integration into existing Monte Carlo/PDE engines in a pricing library. Similar to the local volatility in the diffusion term, one can simply use a new spot-dependent drift to advance the Monte Carlo path or propagate the PDE. It retains the computational cost of the original numerical method but allows one to capture FX and Equity skew so that European Quanto options can be consistently valued.

\section{Numerical Results}
\label{sec:NumericalResults}
We illustrate the accuracy of the new approach with test results where several alternative benchmark methods were used to ensure the testing is comprehensive:
\begin{itemize}
\item 2-D PDE solver;
\item Monte Carlo with $2^{20}$ ($\sim 1$m) simulated Local Vol paths;
\item 2-D integration with Gaussian copula;
\end{itemize}
We calibrate these models to have the same quanto forwards so that the comparison is like-for-like. In addition, similar to \cite{Jackel} we translate the results into implied vol terms so that the it's easier to interpret. The Ad-hoc method samples the At-the-Money vols from the Equity and FX vol surfaces and multiply them with the Equity-FX correlation to get the constant drift adjustment. The functions $\tilde{g}_{i}$ and the CDF functions were obtained from parametrised implied volatility surface fitted to the market with arbitrage removal techniques applied to ensure the tails are well behaved.

\begin{figure}[!htb]
\centering
\begin{subfigure}{0.5\textwidth}
    \centering
    \includegraphics[width=3.0in]{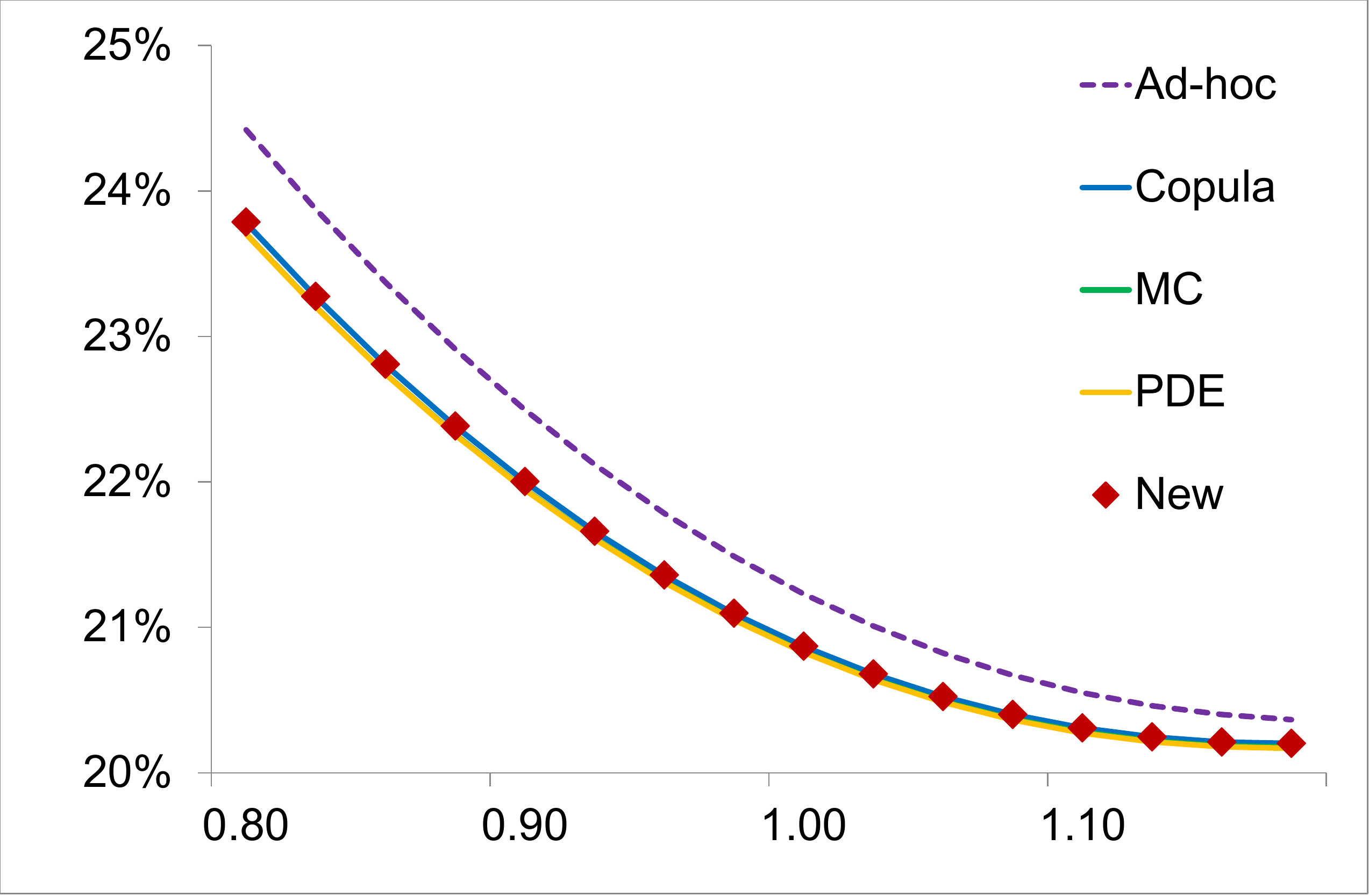}
\end{subfigure}
\captionsetup{singlelinecheck=off}
\caption[.]{Results: .N225 USD 2Y Quanto, Q1 2017 (in implied vol)}
\label{fig:N225USD}
\end{figure}

\begin{figure}[!htb]
\centering
\begin{subfigure}{0.5\textwidth}
    \centering
    \includegraphics[width=3.0in]{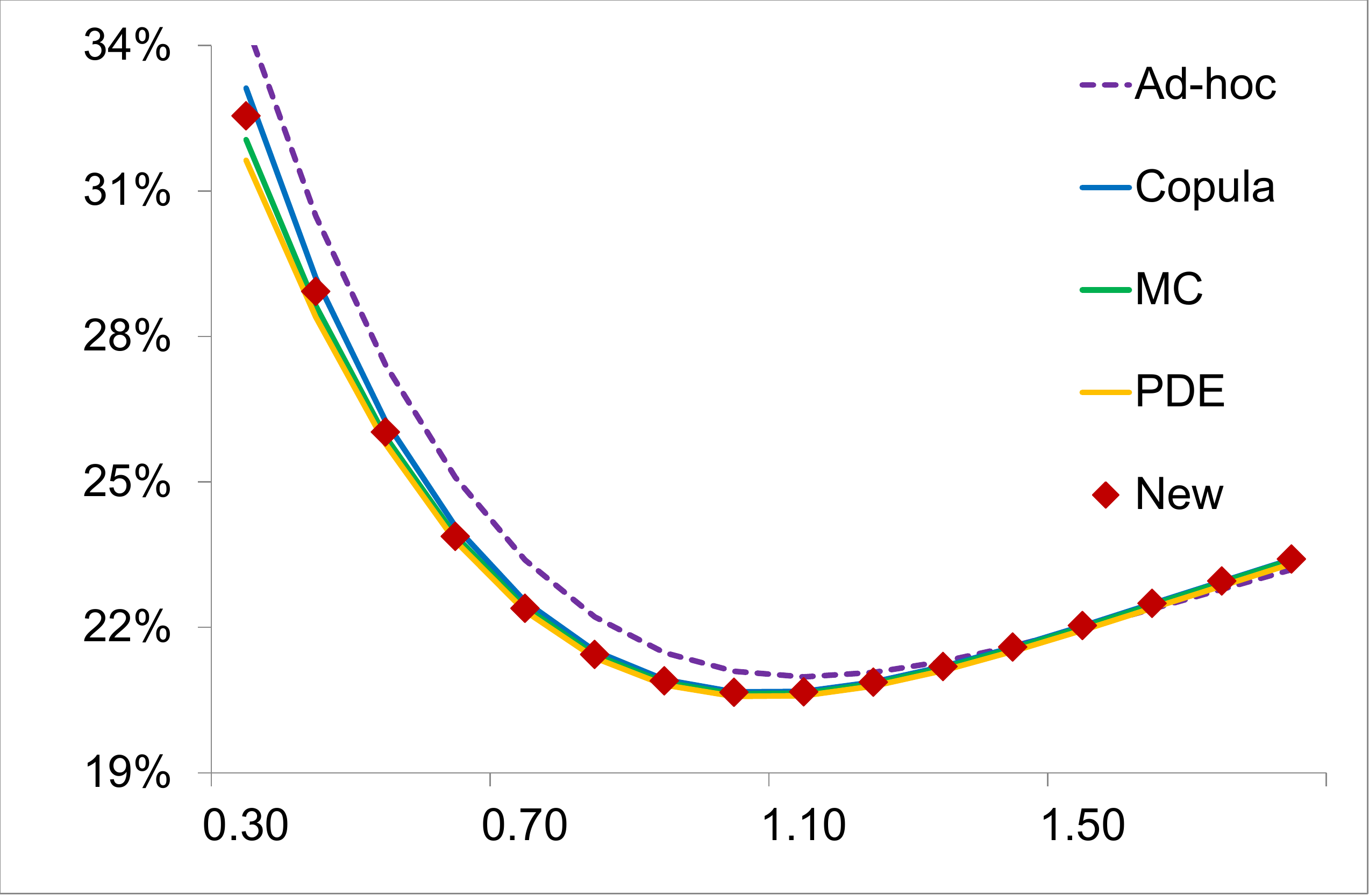}
\end{subfigure}
\captionsetup{singlelinecheck=off}
\caption[.]{Results: .N225 AUD 5Y Quanto, Q1 2017 (in implied vol)}
\label{fig:N225AUD}
\end{figure}

Starting with the USD quanto 2 year option on Nikkei using market data as of January 2017 and comparing the alternative approaches in Figure \ref{fig:N225USD}, we see that the ad-hoc adjustment is over pricing the vols across all strikes versus the other methods, even if the quanto forwards match between all of them. The effect comes from the positive correlation between Nikkei and USDJPY, which remains strongly positive since the financial crisis (around 70\% in the example above) as well as the volatility skew. It is clear that our method shows excellent agreement with the other three standard (slower) benchmark methods. Similar pattern is observed for an equally popular pair, Nikkei quantoed into Australian dollar, where Figure \ref{fig:N225AUD} showing our method remains accurate for extended maturity and strikes.

\begin{figure}[!htb]
\centering
\begin{subfigure}{0.5\textwidth}
    \centering
    \includegraphics[width=3.0in]{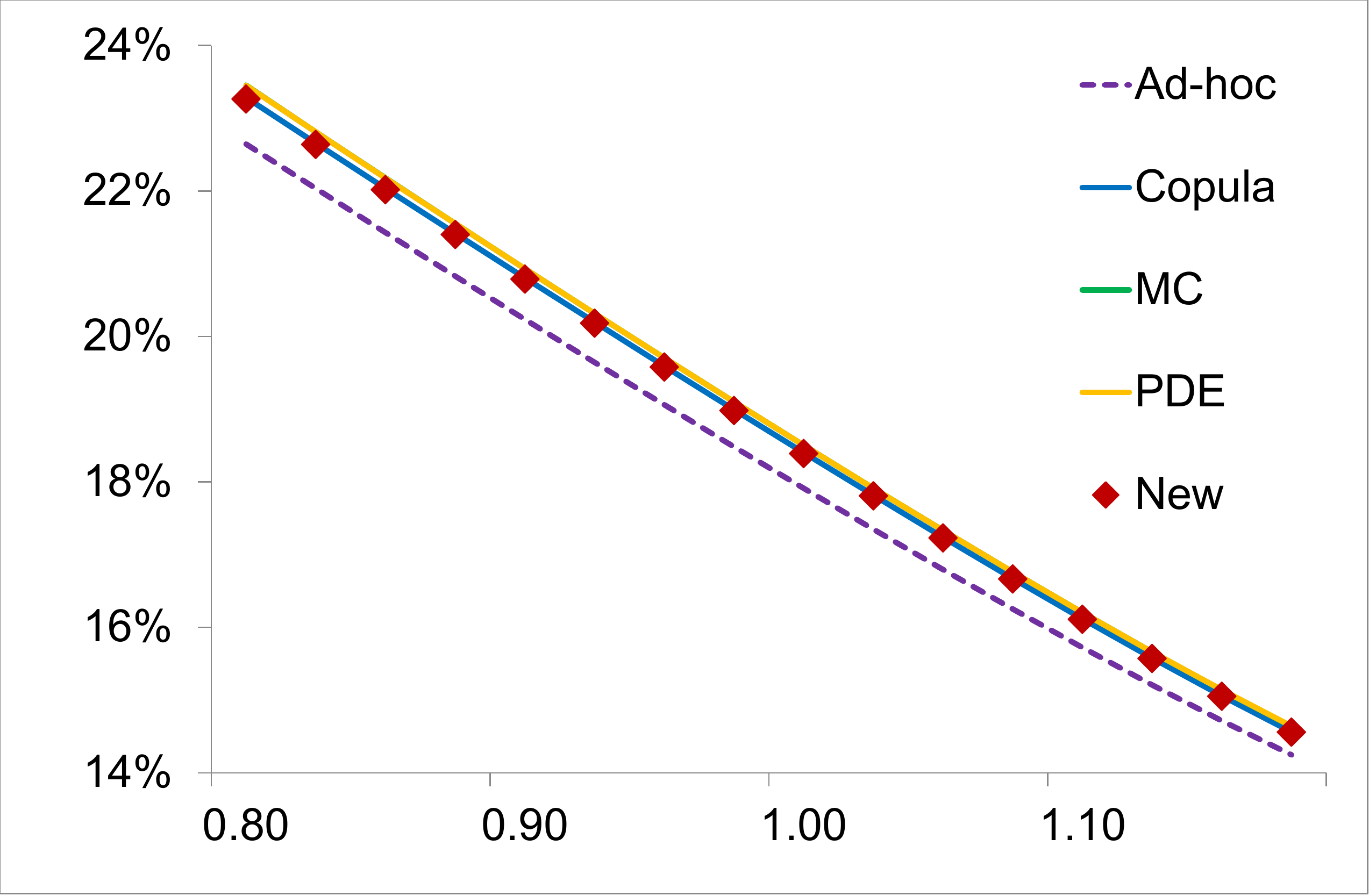}
\end{subfigure}%
\captionsetup{singlelinecheck=off}
\caption[.]{Results: .SPX EUR 2Y Quanto, Q1 2017 ($\rho = -80\% $)}
\label{fig:SPXEUR}
\end{figure}

The Quanto skew effect is directly proportional to the Equity-FX correlation and some of the major index-currency pairs can easily go through regimes of positive, negative or near-zero realized correlation as influenced by supply-demand or political events. We therefore examine the effect of changing signs of correlation by testing S\&P500 index quantoed into EUR and stressing the correlation to be highly negative (-80\%) in Figure \ref{fig:SPXEUR}. As we flip the sign, the ad-hoc bias changes direction as expected while our method retains its close match with other methods, confirming its robustness.

\begin{figure}[!htb]
\centering

\begin{subfigure}{0.5\textwidth}
    \centering
    \includegraphics[width=3.0in]{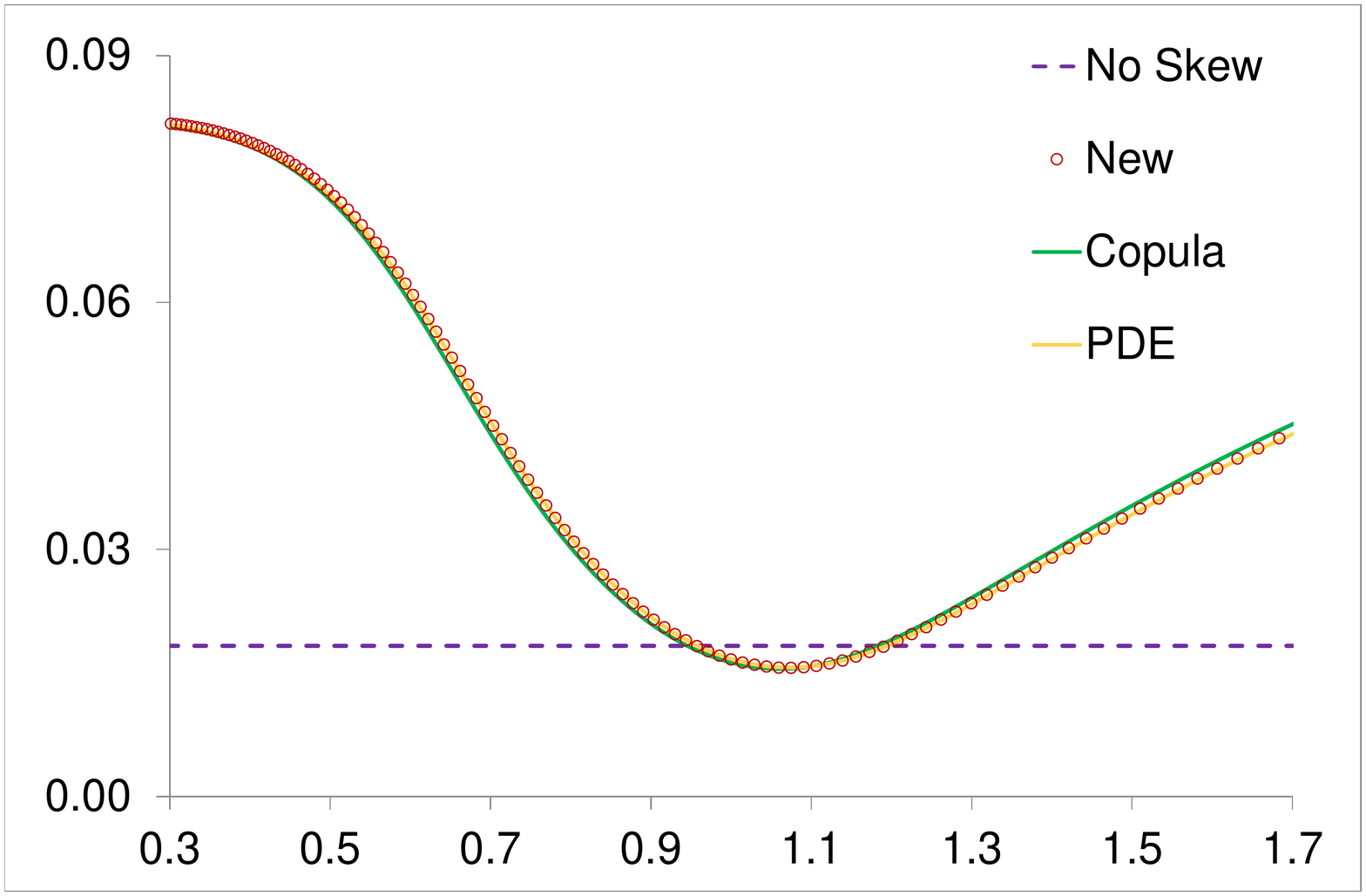}
    \caption{Quanto Local Drift at $t=$6 Month}
\end{subfigure}%
\begin{subfigure}{0.5\textwidth}
    \centering
    \includegraphics[width=3.0in]{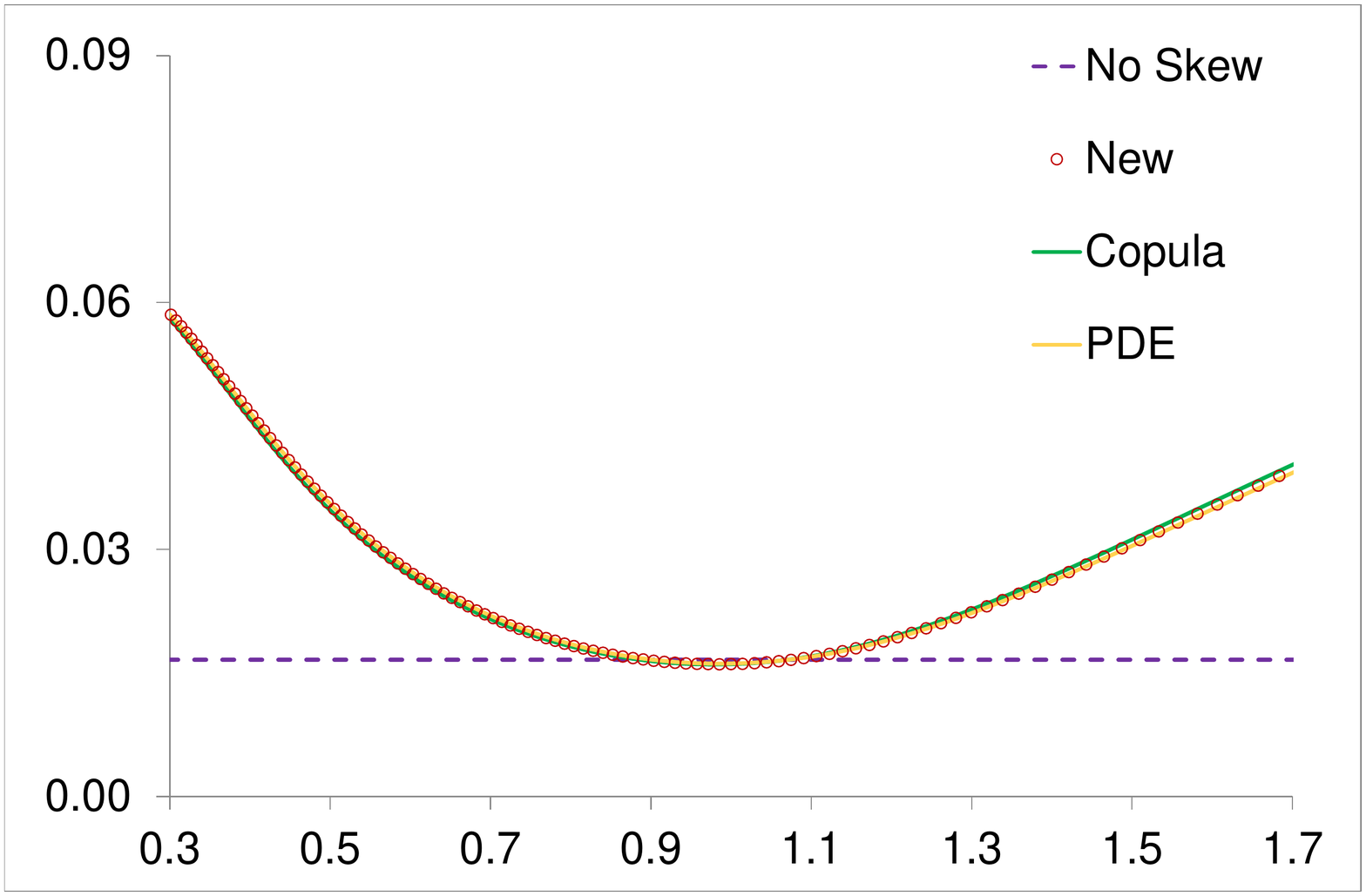}
    \caption{Quanto Local Drift at $t=$2 Year}
\end{subfigure}

\captionsetup{singlelinecheck=off}
\caption[.]{Local Drift Adjustment: Conditional expectation of Quanto Covariance $ \m E_{\m D} \big [ \rho \cdot \sigma_{S}(S, t) \cdot \sigma_{X}(X_t, t) \big | S_t = S \big ]  $ as a function of $ S $}
\label{fig:LocalDrift}
\end{figure}

The final numerical result in Figure \ref{fig:LocalDrift} demonstrates our local drift approximation. Using the same market data for Nikkei and USDJPY, we compute the slice of conditional local volatility function of the FX underlying at $t=$ 6 months and 2 years. Similarly, we demonstrate close match between our very fast approach with the other two benchmark methods: PDE and Copula integration. In contrast to the no-skew case assuming a constant drift derived from the ad-hoc approach, the non-trivial shape of the function is accurately captured by our polynomial.

Finally we report the speed of the proposed method for the Quanto Vanilla option valuation. Implementing the proposed pricing recipe in Section \ref{subsec:QuantoVanillaSpread} in C++ and running the tests with CPU Intel Core i7-4770 3.4 GHz and 32 GB RAM, the average compute time is shown in Table \ref{tab:computeTime}. We report in the last three rows the results of combining hundreds of options in one calculation.

\begin{table}[htbp]
  \centering
    \begin{tabular}{cccll}
    \toprule
    no. of & no. of & no. of & Total time & Per-option \\
    options & maturities & strikes & (seconds) & (seconds) \\
    \midrule
    1 & 1 & 1 &          0.000536119 &  0.000536119 \\
    100 & 1 & 100 &      0.000919493 &  0.000009194 \\
    100 & 10 & 10 &      0.005316215 &  0.000053162 \\
    1000 & 10 & 100 &    0.008435730 &  0.000008436 \\
    \bottomrule
    \end{tabular}
  \caption{Compute time for Quanto Vanilla option spread using proposed pricing recipe}
  \label{tab:computeTime}
\end{table}

Pricing a single option takes only about half a millisecond. This extremely fast performance gets even better if one computes options with the same maturity (and different strikes) simultaneously by only updating the coefficients needed: less than 1 millisecond to compute 100 options (in row 2) and hence a hundredth of that per option. This clearly beats all the other benchmark numerical methods mentioned above by orders of magnitude.

In closing we note that while the method delivers high accuracy and speed in the local volatility/Gaussian copula setting, less precision is expected in stochastic volatility models as observed in \cite{Cesarini}. This is because the \emph{joint} distribution of the two drivers post the marginal quantile transforms in stochastic volatility models deviates from the Gaussian assumption. Potential extension of our approach to address this issue is left as future research.

\section{Conclusion}
\label{sec:Conclusion}
In this article, a new approach of valuing quanto derivatives analytically has been developed, with the aims of high performance and low implementation effort in mind, by applying the stochastic collocation methods introduced by \cite{Grzelaka, Grzelaka2}. Our hope is that the approach is simple enough for participants across different segments of the market and asset classes to start incorporating quanto skew into their pricing and risk management decisions, beyond the use of simplistic constant drift adjustments. This includes trading desks, for which fast calculation is critical, as well as hedge funds and banks, which lean towards simple solutions free from the heavy machinery of sophisticated numerical methods and highly specialised model assumptions.

\section*{Acknowledgement}
The author would like to thank his team members and fellow quants for stimulating discussions on the subject and is particularly grateful to Yonglan Zhu, David Wilkinson and Ghislain Vong.

The views expressed in this article are those of the author alone and do not necessarily represent those of Credit Suisse Group. All errors are the author's responsibility.

\end{document}